\documentclass[prl,twocolumn,nofootinbib]{revtex4-2}
\usepackage{amsmath,amssymb,amsfonts}
\usepackage{graphicx,color}
\usepackage[colorlinks=true, linkcolor=blue, citecolor=blue, linktoc=all]{hyperref}
\usepackage[usenames,dvipsnames]{xcolor}
\usepackage{sfmath}
\usepackage{pict2e}
\makeatletter
\DeclareRobustCommand{\loplus}{\mathbin{\mathpalette\dog@lsemi{+}}}
\DeclareRobustCommand{\lotimes}{\mathbin{\mathpalette\dog@lsemi{\times}}}
\DeclareRobustCommand{\roplus}{\mathbin{\mathpalette\dog@rsemi{+}}}
\DeclareRobustCommand{\rotimes}{\mathbin{\mathpalette\dog@rsemi{\times}}}
\newcommand{\fLie}{\mathfrak{L}}
\newcommand{\dog@rsemi}[2]{\dog@semi{#1}{#2}{-90,90}}
\newcommand{\dog@lsemi}[2]{\dog@semi{#1}{#2}{270,90}}
\newcommand{\dog@semi}[3]{%
  \begingroup
  \sbox\z@{$\m@th#1#2$}%
  \setlength{\unitlength}{\dimexpr\ht\z@+\dp\z@\relax}%
  \makebox[\wd\z@]{\raisebox{-\dp\z@}{%
    \begin{picture}(1,1)
    \linethickness{\variable@rule{#1}}
    \roundcap
    \put(0.5,0.5){\makebox(0,0){\raisebox{\dp\z@}{$\m@th#1#2$}}}
    \put(0.5,0.5){\arc[#3]{0.5}}
    \end{picture}%
  }}%
  \endgroup
}
\newcommand{\variable@rule}[1]{%
  \fontdimen8  
  \ifx#1\displaystyle\textfont3\else
    \ifx#1\textstyle\textfont3\else
      \ifx#1\scriptstyle\scriptfont3\else
        \scriptscriptfont3\relax
  \fi\fi\fi
}
\begin{document}
\newcommand{\thistitle}{Embeddings and Integrable Charges for Extended Corner Symmetry}
\newcommand{\ulb}{Physique Math\'ematique des Interactions Fondamentales \& International Solvay Institutes, Universit\'e Libre de Bruxelles, Campus Plaine - CP 231, 1050 Bruxelles, Belgium}
\newcommand{\uiuc}{Illinois Center for Advanced Studies of the Universe \& Department of Physics, University of Illinois, 1110 West Green St., Urbana IL 61801, U.S.A.}

\newcommand{\brac}[2]{\{\kern-2pt[ #1,#2]\kern-2pt\}}
\newcommand{\un}[1]{\underline{#1}}
\newcommand{\pa}{\partial}
\newcommand{\RR}{\mathbb{R}}
\newcommand{\Luca}[1]{{\color{red}\textbf{LC: \ #1}}}
\newcommand{\ack}[1]{[{\bf Pfft!: {#1}}]}
\newcommand{\beq}{\begin{eqnarray}}
\newcommand{\eeq}{\end{eqnarray}}
\newcommand{\beqn}{\begin{eqnarray}}
\newcommand{\eeqn}{\end{eqnarray}}
\newcommand{\hlt}[1]{{\color{WildStrawberry}{\em #1}}\index{#1}}
\newcommand{\dM}{\mathfrak{diff}(M)}
\newcommand{\dS}{\mathfrak{diff}(S)}

\title{\thistitle}
\author{Luca Ciambelli}
\affiliation{\ulb}
\author{Robert G. Leigh}
\author{Pin-Chun Pai}
\affiliation{\uiuc}
\begin{abstract}
We revisit the problem of extending the phase space of diffeomorphism-invariant theories to account for embeddings associated with the boundary of sub-regions. We do so by emphasizing the importance of a careful treatment of embeddings in all aspects of the covariant phase space formalism. In so doing we introduce a new notion of the extension of field space associated with the embeddings which has the important feature that the Noether charges associated with all extended corner symmetries are in fact integrable, but not necessarily conserved. We give an intuitive understanding of this description. We then show that the charges give a representation of the extended corner symmetry via the Poisson bracket, without central extension.
\end{abstract}
\maketitle

\section{Introduction}

Classical gravitational theories are gauge theories for which the principal symmetry is diffeomorphism invariance.
The study of symmetry in gauge theories has a long history going back to the seminal work of Noether \cite{Noether1918}.
While the first Noether theorem deals with global symmetries, Noether herself remarked that for a gauge symmetry, defined as a local symmetry acting differently at different spacetime points, the first theorem gives a current associated to the symmetry that necessarily vanishes on-shell up to total derivatives. It then follows that Noether charges for gauge symmetries must be defined as surface integrals of $d-2$ forms, where $d$ is the spacetime dimension, i.e., as surface charges. Thus codimension-2 surfaces play a central role in gravity and other gauge theories. We will refer to them here as {\it corners} \cite{Donnelly:2016auv,Speranza:2017gxd,Freidel:2020ayo,Freidel:2020svx,Freidel:2020xyx,Ciambelli:2021vnn,Freidel:2021cjp} for brevity.

The first discussion of the role of gauge symmetries and their surface charges appeared using the Hamiltonian formalism in the seminal work of Regge and Teitelboim \cite{Regge:1974zd}. Later, Wald and others \cite{Wald:1993nt,Iyer:1994ys,Wald:1999wa}, using the covariant phase-space formalism, and  Barnich and Brandt \cite{Barnich:2001jy}, using Anderson's variational bicomplex \cite{MR1188434, Anderson:1996sc}, formulated the Lagrangian analysis of gauge symmetries and their conserved quantities. The theory of asymptotic symmetries and surface charges following from these works has been shown to give the same results, modulo ambiguities, as reviewed for example in \cite{Compere:2019qed}.

In any diffeomorphism-invariant classical theory on a manifold $M$, in the absence of any extra structure we are free to perform any diffeomorphism, as it corresponds to a gauge redundancy. However, in the presence of some geometric structure, such as a subregion of $M$ or a corner,
some diffeomorphisms will become physical symmetries. This is a familiar feature of other gauge theories as well, where in the corresponding quantum theory, non-trivial physics is involved in the gluing of subregions and is implicated in entanglement properties. Often one addresses such geometric structures through the mathematical construction of embeddings \cite{Donnelly:2016auv,Speranza:2017gxd,Ciambelli:2021vnn}. 
In particular, in \cite{Ciambelli:2021vnn} we explored the consequences of carefully treating such embeddings, and found a theory-independent finite 
sub-algebra of $\dM$, compatible with the presence of an embedded submanifold $S$, whose Lie brackets close on itself, which we referred to as the maximal embedding symmetry, ${\cal A}_k=(\dS\loplus \mathfrak{gl}(k,\RR))\loplus\RR^k$.  Applied to corners ($k=2$), this algebra includes the so-called extended corner algebra \cite{Freidel:2021cjp}, which in turn includes notable sub-cases, such as the BMSW algebra \cite{Freidel:2021fxf}, the generalized BMS algebra \cite{Campiglia:2014yka,Compere:2018ylh,Campiglia:2020qvc}, the extended BMS algebra \cite{Barnich:2010eb,Barnich:2011mi}, as well as the original BMS algebra \cite{doi:10.1098/rspa.1962.0161,doi:10.1098/rspa.1962.0206,Sachs:1962wk}. 

In diffeomorphism-invariant theories, a basic construction is the integration of a Noether charge density over  a corner. An important issue in this regard is {\it integrability}, whether or not a diffeomorphism is equivalently associated with a Hamiltonian vector field on the space of fields of the theory and a function on field space. If integrability pertains, then diffeomorphisms are represented by such functions and the algebra of charges reproduces canonically the vector field algebra via the Poisson bracket, possibly with central extension \cite{Brown:1986nw}. A separate issue is how the possible non-conservation of charge in a given region is associated with the flow of flux in or out of the region. 
As advocated in \cite{Donnelly:2016auv}, such features can be obtained in general if one extends the field space of the theory to include effects associated with embedded submanifolds. 
However, in that and subsequent analyses, it was found that diffeomorphisms that do not preserve a corner are on a much different footing than those that do: only the latter were found to correspond to integrable charges. The non-trivial diffeomorphisms that do not preserve a corner are its normal translations, which appear in the $\RR^2$ factor in the extended corner symmetry ${\cal A}_2=(\dS\loplus \mathfrak{gl}(2,\RR))\loplus\RR^2$. In this paper we are primarily interested in generic corners that are at finite distance, but we note that in the context of asymptotic symmetries, it is the $\RR^2$ factor that gives rise to the so-called supertranslations, and so it is certainly of central interest to fully understand it. 

In \cite{Ciambelli:2021vnn} by carefully treating embeddings we found that the maximal embedding algebra corresponds precisely to the set of those diffeomorphisms that become physical in the presence of a corner and furthermore that the variation of the charge with respect to any diffeomorphism gives precisely $\delta_{\un\eta}H_{\un\xi}=H_{[\un\xi,\un\eta]}$, which appears ready-made for an interpretation in terms of Poisson brackets. This result was obtained purely geometrically without reference to the covariant phase space formalism beyond a specification of the Noether charge density.  

In this letter, we will show how to extend the field space of diffeomorphism-invariant theories such that the entire extended corner symmetry is realized by integrable charges. This is achieved by introducing the same careful treatment of embeddings into the covariant phase space methodology, which leads to a natural extension of the symplectic structure of the theory which differs from that introduced in \cite{Donnelly:2016auv}. The intuitive picture is that one can accommodate symmetries that ``move the corner'' by systematically keeping track of variations of the embedding map.
We stress that integrability is achieved on the entire field space (on-shell) without a need for first setting some flux to zero. The existing literature contains many examples where integrability is achieved only in this sense. 
Here we propose a universal resolution and the consequences are far-reaching:  we find that the algebra of charges associated with the full extended corner symmetry (and thus any of its subalgebras appropriate to any particular physical situation) is represented in terms of Poisson brackets on the extended phase space, without central extension. 
As a further consequence, the flux is accounted for fully by the field-space contraction of the Hamiltonian vector field associated to the symmetry and the extended pre-symplectic potential.


\section{Embedding Maps}

In this section, we describe in detail embedding maps. This material of course is standard but it is important to carefully establish our notation. We are concerned entirely here with classical diffeomorphism-invariant theories. We thus suppose that we have a spacetime manifold $M$ of dimension $d$ that we endow locally with coordinates via the trivialization $\Phi_i:U_i\to \RR^d, U_i\subset M$, giving $\Phi_i:p\mapsto y^\mu$. An embedded submanifold of $M$ of codimension $k$ is the image of a (smooth)  map  $\phi_k:S_k\to M$, where $S_k$ is a manifold of dimension $d-k$. We introduce coordinates $\{\sigma^\alpha\}$ on $S_k$ with the embedding described by $\phi_k:\{\sigma^\alpha\}\mapsto \{y^\mu(\sigma)\}$, the latter thought of as the locus of points of the embedded manifold in $M$. Given the embedding, differential forms $\alpha\in\Gamma(\wedge^{d-k}T^*M)$ can be pulled back to a top form $\phi_k^*\alpha\in\Gamma(\wedge^{d-k}T^*S_k)$ on $S_k$. Such a form can then be integrated over $S_k$. 

We will be interested in describing a variation of such an embedding $\phi_k\to\phi_k+\delta\phi_k$, promoted to an arbitrary variation from the perspective of covariant phase space. Given such, we can then express the response of the embedding to a diffeomorphism as a contraction\footnote{Throughout the paper, we use the notation $i_{\un\xi}$ for the contraction of a spacetime vector field with a differential form and $I_{V_{\un\xi}}$ for the corresponding contraction on field space; here $V_{\un\xi}$ denotes the vector field on field space associated with the spacetime diffeomorphism generated by $\un\xi$. Similarly, Lie derivatives are denoted ${\cal L}_{\un\xi}=di_{\un\xi}+i_{\un\xi}d$ and $\fLie_{V_{\un\xi}}=\delta I_{V_{\un\xi}}+I_{V_{\un\xi}}\delta$.} of $\delta\phi_k$ with a vector field on phase space, i.e., $\delta_{\un\eta}\phi_k = I_{V_{\un\eta}}\delta\phi_k$. 
Essentially, this idea was implemented originally in \cite{Donnelly:2016auv} by writing the corresponding change of the embedding coordinates as
\beq\label{varyembk}
 \{\sigma^\alpha\}\rightarrow \{y^\mu(\sigma^\alpha)+{{\chi}}^\mu(\sigma^\alpha)\}
\eeq
which we  interpret as defining a vector field on $M$, ${\underline{\chi}} \in TM$ defined at all points on the embedded manifold. Thus we have $\delta\phi_k \leftrightarrow \un\chi$ and we will consequently regard $\un\chi$ as a 1-form on field space, the details of which we will review below.

Consider a $(d-k)$-form $\alpha[g,y]$ where $g$ collectively denotes the fields of a theory and $y$ denotes the coordinates on $M$ in which we are working. Given 
an embedding $\phi_k:S_k\to M$, we can pull back $\alpha$ to $S_k$ and integrate
\beq\label{genericform}
A[g,\phi] := \int_{S_k} \phi_k^*\Big(\alpha[g,y]\Big).
\eeq
Often in the literature these details are not expressed clearly and instead integrals are expressed directly in the coordinates of $M$. We will see that keeping track of such details will have important consequences for the covariant phase space formalism. An important result that we will use repeatedly is Stokes' theorem, which in the notation we are using reads
\beq
\int_{S_k}\phi^*_k\Big(d\beta[g,y]\Big)=\int_{\pa S_k}\phi^*_{k-1}\Big(\beta[g,y]\Big)
\eeq
where $\pa S_k$ is the boundary of $S_k$, its embedding given by $\phi_{k-1}:\pa S_k\to M$.

In \cite{Ciambelli:2021vnn}, we made use of this formalism in a simple setting, in which we simply integrated the Noether charge density in this way over a corner, corresponding to the case $k=2$, 
\beq\label{firstCH}
H_{\un\xi}=\int_{S_2} \phi_2^*(Q_{\un\xi})=\int_{S_2} \phi_2^*(*di_{\un\xi}g),
\eeq
where $\un\xi\in TM$ is an arbitrary vector field on $M$ and $Q_{\un\xi}$ is the Noether charge density, making no further reference to any other aspects of the covariant phase space formalism. In the second equality we inserted the specific representation of the  Noether charge for the Einstein-Hilbert theory. We found that only those vector fields in the maximal embedding symmetry\footnote{The tilde denotes the fact that whereas the maximal embedding algebra is the algebra of the group $\Big(Diff(S)\ltimes GL(k,\RR)\Big)\ltimes \RR^k$, only the group $\Big(Diff(S)\ltimes SL(k,\RR)\Big)\ltimes \RR^k$ is realized in the Einstein-Hilbert theory.} $\tilde{\cal A}_k=\Big(\mathfrak{diff}(S)\loplus \mathfrak{sl}(k,\RR)\Big)\loplus \RR^k$ give rise to non-zero charges. Thus all other diffeomorphisms remain pure gauge, even in the presence of a corner. We then showed 
that
\beq
\delta_{\un\xi}H_{\un\eta} 
=-H_{[\un\xi,\un\eta]}.
\eeq
It is tempting to interpret this result as a bracket
\beq
\brac{ H_{\un\xi}}{H_{\un\eta}}=-H_{[\un\xi,\un\eta]}.
\eeq
If this bracket were to coincide with the Poisson bracket of the covariant phase space formalism for all $\tilde{\cal A}_2$ diffeomorphisms, then we would conclude that the symmetry is represented on phase space without central extension. 

We will show in the remainder of the paper that precisely this pertains if we correctly enlarge the field space to include the degrees of freedom associated with embedding maps corresponding to subregions of $M$ and hypersurfaces, and extend the symplectic structure accordingly.  Such a correction is in the spirit of \cite{Donnelly:2016auv} but differs in a crucial way. In particular we will show that the correct extension of the symplectic structure yields integrable charges for {\it all} of $\tilde{\cal A}_2$  and furthermore, non-conservation of the charges associated with  flux is fully accounted for by the extended symplectic structure.

Consider the variation of \eqref{genericform}
\beq
\delta A[g,\phi]=\delta \int_{S_k} \phi_k^*(\alpha[g,y]).
\eeq
 The variation receives two contributions, one from the variation of $\alpha$, and one from the variation of the embedding $\phi_k$,
\beq
\delta A[g,\phi]= \int_{S_k} \phi_k^*\Big(\delta\alpha[g,y]\Big)+\int_{S_k} \delta\phi_k^*\Big(\alpha[g,y]\Big)
\eeq
where as usual
\begin{eqnarray}
\label{variation2}
\delta \alpha[g,y] := \alpha[g+\delta g,y] - \alpha[g,y]
\end{eqnarray}
while the variation of the embedding is given by  \begin{eqnarray}
\label{variation3}
(\delta \phi_k^*) (\alpha[g,y]) &:=& \phi_k^*\Big(\alpha[g,y+\chi] - \alpha[g,y]\Big)\\
&=&\phi_k^*\Big( {\cal L}_{\un\chi} \alpha[g,y] \Big)
\end{eqnarray}
where we are using the notation introduced in eq. \eqref{varyembk}.
Thus
\beq\label{variationA}
\delta A[g,\phi]= \int_{S_k} \phi_k^*\Big(\delta\alpha[g,y]+{\cal L}_{\un\chi} \alpha[g,y]\Big).
\eeq
The variation with respect to a diffeomorphism in particular is given by
\beq
I_{V_{\un\eta}}\delta g = \fLie_{V_{\un\eta}}g = {\cal L}_{\un\eta}g,\qquad
I_{V_{\un\eta}}\un\chi=-\un\eta\Big|_{\phi_k(S_k)}
\eeq
which implies in particular that $\delta_{\un\eta}\phi_k = I_{V_{\un\eta}}\delta\phi_k$.
Essentially similar relations have also appeared in  \cite{Donnelly:2016auv,Speranza:2019hkr} for example. 
Before moving on to covariant phase space formalism, we consider the implication of the nilpotency of $\delta$,
\beqn
0=\delta^2A[g,\phi]=\int_{S_k}\phi_k^*\Big({\cal L}_{\delta\underline{\chi}}\alpha[g,y]+\tfrac12 {\cal L}_{[\underline{\chi},\underline{\chi}]}\alpha[g,y]\Big)\nonumber
\eeqn
from which we conclude 
\beq\label{delta_chi}
\delta \un\chi = -\tfrac12[\un\chi,\un\chi].
\eeq
Thus $\un\chi$ is not an exact form in field space generally. We note that this result resembles a property of  ghosts, which is by no means an accident. Indeed this is another sign that in the presence of an embedding, a `vertical` degree of freedom becomes physical. Related ideas appear in \cite{Gomes:2016mwl,Speranza:2017gxd}.

\section{Covariant Phase Space}

In the case where one integrates a Lagrangian density over all of the manifold $M$, no embedding is required because the Lagrangian is a top-form on $M$. However, suppose we consider integrating the Lagrangian over some subregion $R$ of $M$; we will regard that subregion as embedded in $M$ via $\phi_0:R\to M$, with the boundary of the subregion  regarded as an embedded hypersurface in $M$. So we write 
\beq
S_{R}[g,\phi]=\int_R \phi^*_0\Big(L[g,y]\Big).
\eeq
Given this, we then write
\beqn
\delta S_{R}[g,\phi]&=& \int_{R} \phi_{0}^*\Big(\delta L[g,y] \Big) 
+   \int_{R} \delta\phi_{0}^*\Big( L[g,y]\Big)
\nonumber\\
&\hat{=}& \int_{ \pa R} \phi_{1}^* \Big( \theta[g,\delta g, y]\Big)
+   \int_{R} \phi_{0}^* \Big({\cal L}_{\un\chi} L[g,y] \Big)
\nonumber\\
&\hat{=}&
\int_{\pa R} \phi_1^*\Big(\theta[g,\delta g,y]+i_{\un\chi}L[g,y]\Big).
\eeqn
where $\theta$ is the usual pre-symplectic form satisfying $\delta L\,\hat{=}\,d\theta$. 
This result suggests a  corresponding extended pre-symplectic potential,
\beq
\label{extended_potential}
\Theta^{ext.}_\Sigma  \equiv \int_{\Sigma} \phi^*_1\left (\theta[g,\delta g,y]+i_{\un\chi}L[g,y] \right)
\eeq
where $\phi_1$ now embeds a hypersurface $\phi_1:\Sigma\to M$, possibly with boundary. This extension is different than the one first introduced in \cite{Donnelly:2016auv}. To understand its properties, we consider the corresponding extended pre-symplectic structure, which can be shown to be of the form
\beqn\label{OmegaExt}
\Omega^{ext.}_\Sigma &:=& \delta \Theta^{ext.}_\Sigma 
=\int_{\Sigma} \phi^*_1\Big(\delta\theta[g,\delta g,y]\Big)
\\
&&+\int_{\pa\Sigma}\phi^*_2\Big(i_{\un\chi}\theta[g,\delta g,y]+\tfrac12i_{\un\chi}i_{\un\chi}L[g,y] \Big).\nonumber
\eeqn
We then identify 
\beq
\Omega^{ext.}_\Sigma[g,\delta g,\un\chi,y] = \Omega_\Sigma[g,\delta g,y]+\Omega^{cor.}_{\pa\Sigma}[g,\delta g,\un\chi,y].
\nonumber\eeq
Note that if embeddings are not considered as in the standard treatment, then the result would reduce to the first line of \eqref{OmegaExt}. This is only consistent however if we restrict attention to diffeomorphisms that preserve the corner $\pa\Sigma$.
On the other hand,
similar extended structures were given in \cite{Donnelly:2016auv,Speranza:2017gxd} but the prescription used there differs from ours in an important way, the details of which we will discuss below. 

We now turn our attention to the question of integrability, and so we compute
\beq
I_{V_{\un\eta}}\Omega^{ext.}_\Sigma =
I_{V_{\un\eta}}\Omega_\Sigma+I_{V_{\un\eta}}\Omega^{cor.}_{\pa\Sigma}.
\nonumber\eeq
For the first term, we have the standard result
\beqn
I_{V_{\un\eta}}\Omega_\Sigma=\int_\Sigma \phi_1^*\Big(I_{V_{\un\eta}}\delta\theta\Big)
\hat{=}\int_\Sigma \phi_1^*\Big(di_{\un\eta}\theta
-\delta dQ_{\un\eta}\Big)
\eeqn
where here we have assumed that $\un\eta$ is an arbitrary field-independent vector field as well as the covariance of $\theta$,  $\fLie_{V_{\un\eta}}\theta={\cal L}_{\un\eta}\theta$ \cite{Iyer:1994ys} (see also \cite{Chandrasekaran:2020wwn}). 
Here $Q_{\un\eta}$ is the Noether charge density, defined as usual by
\beq
dQ_{\un\eta}=I_{V_{\un\eta}}\theta-i_{\un\eta}L.
\eeq
Next, we consider the corner contribution,
\beqn
I_{V_{\un\eta}}\Omega^{cor.}_\Sigma
&=&
\int_{\pa\Sigma}\phi^*_2\Big(I_{V_{\un\eta}}i_{\un\chi}\theta+\tfrac12I_{V_{\un\eta}}i_{\un\chi}i_{\un\chi}L \Big)\nonumber
\\
&=&
-\int_{\pa\Sigma}\phi^*_2\Big(i_{\un\eta}\theta
+i_{\un\chi}I_{V_{\un\eta}}\theta
+\tfrac12i_{\un\eta}i_{\un\chi}L 
-\tfrac12i_{\un\chi}i_{\un\eta}L
\Big)\nonumber
\\
&=&
-\int_{\pa\Sigma}\phi^*_2\Big(i_{\un\eta}\theta
+i_{\un\chi}(dQ_{\un\eta}+\tfrac12i_{\un\eta}L )
+\tfrac12i_{\un\eta}i_{\un\chi}L 
\Big)\nonumber
\\
&=&
-\int_{\pa\Sigma}\phi^*_2\Big(i_{\un\eta}\theta
+i_{\un\chi}dQ_{\un\eta} 
\Big)
\eeqn
and so
\beqn
I_{V_{\un\eta}}\Omega^{ext.}_\Sigma\hat{=}
-\int_{\Sigma}\phi^*_1\Big(
\delta d Q_{\un\eta}
+{\cal L}_{\un\chi}dQ_{\un\eta}\Big)
=-\delta\int_{\pa\Sigma}\phi^*_2\Big(
 Q_{\un\eta}\nonumber
\Big)
\eeqn
where we made use of eq. \eqref{variationA} in reverse. Thus the charge is integrable, and we write
\beqn
I_{V_{\un\eta}}\Omega^{ext.}_\Sigma
=-\delta\int_{\pa\Sigma}\phi^*_2
 (Q_{\un\eta})=-\delta H_{\un\eta}
\eeqn
Thus $V_{\un\eta}$ is the Hamiltonian vector field on field space corresponding to the Hamiltonian function $H_{\un\eta}$, the latter coinciding precisely with that appearing in \eqref{firstCH}. Continuing in the same vein, it can be shown that
\beqn
I_{V_{\un\xi}}I_{V_{\un\eta}}\Omega^{ext.}_\Sigma
=-I_{V_{\un\xi}}\delta\int_{\pa\Sigma}\phi^*_2
 (Q_{\un\eta})=\int_{\pa\Sigma}\phi^*_2
 (Q_{[\un\xi,\un\eta]})= H_{[\un\xi,\un\eta]}\nonumber
\eeqn
which coincides with the result of \cite{Ciambelli:2021vnn}.
Thus we have the Poisson bracket
\beq
\Big\{H_{\un\xi},H_{\un\eta}\Big\}=\delta_{\un\xi}H_{\un\eta}=-H_{[\un\xi,\un\eta]}.
\eeq
This proves that the charges are canonically represented, given the choice \eqref{extended_potential}. This is achieved without any restriction on the field space, without boundary conditions being imposed and without involvement of ambiguities that would alter the definition of the Noether charge.

Finally we note that $\Theta^{ext.}_\Sigma$ correctly accounts for  flux. Consider for example a `pillbox' with spacelike boundaries $\Sigma_{1,2}$ which have boundaries $S_{1,2}$ and a timelike boundary $B$ (whose boundary is $S_1\cup S_2$). Then
\beqn
H_{\un\xi}\Big|_{S_2}-H_{\un\xi}\Big|_{S_1}&=& 
\int_B \phi^*_1(dQ_{\un\xi})\\
&=& \int_{B}\phi^*_1\Big(I_{V_{\un\xi}}\theta-i_{\un\xi}L\Big)
\\
&=& I_{V_{\un\xi}}\int_{B}\phi^*_1\Big(\theta+i_{\un\chi}L\Big)
\\
&=& I_{V_{\un\xi}}\Theta_B^{ext.}
\eeqn
where we used again our definition \eqref{extended_potential} applied to $B$. Thus the flux of the extended symplectic form precisely accounts for possible non-conservation of the charges. This applies to all vector fields $\un\xi$, whether it be tangential or orthogonal.

\subsection{Comparison to Previous Treatments}

In the usual treatment, where embeddings are not considered at all, one has $\Theta_\Sigma=\int_\Sigma \theta$ and $I_{V_{\un\eta}}\Omega_\Sigma 
=- \int_{\pa\Sigma}\Big(\delta Q_{\un\eta}-i_{\un\eta}\theta\Big)
$ as above. Thus, one says that the charge is integrable if $i_{\un\eta}\theta$ is either zero, or is set to zero by imposing some condition. Note that it is automatically zero if $\un\eta$ is tangent to $\pa\Sigma$, but not if $\un\eta$ is normal. 
In fact, this is precisely the problem that we have resolved by including the embedding -- the charges are then integrable for all $\un\eta$. 

Finally we consider the relationship of our treatment to that of   \cite{Donnelly:2016auv,Speranza:2017gxd}. In fact, they chose a different extension by making use of an ambiguity in the pre-symplectic form
\beqn
\Theta^{DF}_{\Sigma}
&=&\Theta^{ext.}_{\Sigma}+ \int_{\pa\Sigma}\phi_2^*\Big(Q_{\un\chi}\Big)
\\
&=& \int_{\Sigma}\phi_1^*\Big(\theta+i_{\un\chi}L+dQ_{\un\chi}\Big).
\eeqn
This has the property that $I_{V_{\un\eta}}\Theta^{DF}_\Sigma=0$, and thus all diffeomorphism charges have been made to vanish. However, as we have seen, it is  the pre-symplectic structure $\Theta^{ext.}_\Sigma$ that gives rise to the actual Noether charges that generate the extended corner symmetry.

\section{Conclusions}

In this letter, we have emphasized the need for an interpretation in which physical symmetries are associated with corners and we have found the description of field space that gives rise to that interpretation. We plan to follow up this work with a more extensive account in which we study many of the examples of current interest and extend the construction to field-dependent vector fields and the corresponding modified brackets. For example, one may anticipate that our formalism has important implications when applied to the case of black hole horizons as well as asymptotic structures. The latter case in particular requires modifications to account for renormalization, which is currently under investigation.

\paragraph{Acknowledgements} We thank the participants of the First Online Corners Workshop for discussions during and after the presentation of these results. We also thank L. Freidel for further communications.
The research of LC was partially supported by a Marina Solvay 
Fellowship, by the ERC Advanced Grant ``High-Spin-Grav" and by 
FNRS-Belgium (convention FRFC PDR T.1025.14 and convention IISN 4.4503.15).
The work of RGL was supported by the U.S. Department of Energy under contract DE-SC0015655.

%
\end{document}